\documentclass{article}

\usepackage[preprint]{neurips_2026}
\makeatletter
\renewcommand{\@notice}{}
\makeatother

\usepackage[utf8]{inputenc}
\usepackage[T1]{fontenc}
\usepackage[hidelinks]{hyperref}
\usepackage{url}
\usepackage{booktabs}
\usepackage{amsfonts}
\usepackage{amssymb}
\usepackage{amsmath}
\usepackage{amsthm}
\usepackage{nicefrac}
\usepackage{microtype}
\usepackage{graphicx}
\usepackage{tikz}
\usepackage{wrapfig}
\usepackage{algorithm}
\usepackage{algpseudocode}
\usepackage{placeins}
\usetikzlibrary{arrows.meta, positioning, decorations.pathreplacing}

\newtheorem{lemma}{Lemma}

\title{Encrypt What Matters: When Selective Homomorphic Inference Is Efficient}

\author{%
  Ali Backour \quad Juan Reyes \quad Jaime Punyed \quad Ana Onoprishvili \\
  Massachusetts Institute of Technology \\
  Cambridge, MA 02139 \\
  \texttt{\{abackour, jereyes, jspun34, onopre\}@mit.edu}
}

\begin{document}
\maketitle

\begin{abstract}
Fully homomorphic encryption (FHE) enables inference on private data without revealing it to the
server, but evaluating an entire input under FHE is expensive. We study \emph{selective homomorphic
inference}, where only a sensitive region of interest (ROI) is encrypted, and computations independent
of that region are performed in plaintext. Selective evaluation produces the same output as full FHE on the same model, without retraining. Its efficiency
depends on how quickly encrypted dependencies spread through the network. For small encrypted ROIs, locality-preserving architectures can achieve order-of-magnitude homomorphic-evaluation speedups, whereas architectures with early global mixing provide essentially no speedup. These results identify locality as the key architectural property governing the benefit of selective homomorphic inference.
\end{abstract}

\section{Introduction}

Machine-learning inference is increasingly performed by remote services: a client sends an input to
a server that owns a trained model and receives its prediction. This is convenient, but problematic
when the input contains private information such as a medical image, an identity document, or a
personal photograph.
Fully homomorphic encryption (FHE) provides a way to run such inference without revealing the
input to the server~\citep{10.1145/1536414.1536440}. With ordinary encryption, data must normally be decrypted before they can be
processed, while FHE allows computation to be performed directly on encrypted values: the client
encrypts an input $\mathbf{x}$, the server evaluates the model on the ciphertext and returns an
encrypted prediction that only the client can decrypt. The server therefore performs the inference
without seeing the protected data.

The main obstacle is cost. Operations on encrypted values are substantially more expensive than
ordinary arithmetic, making homomorphic neural-network inference much slower than plaintext
inference~\citep{pmlr-v48-gilad-bachrach16, chou2018fastercryptonetsleveragingsparsity}. A large body of work has therefore focused on making
encrypted inference cheaper. CryptoNets first demonstrated neural-network inference on encrypted
inputs~\citep{pmlr-v48-gilad-bachrach16}, and subsequent systems improved its practicality through FHE-friendly
architectures, discretizations, sparsity, packing, and programmable bootstrapping
~\citep{cryptoeprint:2017/1114, brutzkus2019lowlatencyprivacypreserving, chou2018fastercryptonetsleveragingsparsity, cryptoeprint:2021/091}. These approaches largely assume that the
entire input is private and therefore encrypt the complete input.

In many applications, however, sensitivity is spatially local. A photograph may contain a sensitive
face, a scan a particular lesion, or a document a name or identifier, while much of the remaining
input can be public. Recent work has therefore considered encrypting only sensitive regions.
PFHE~\citep{dai2025pfhe} selectively encrypts privacy-sensitive image regions for CNN inference, while
$\Pi_{\mathrm{ROI}}$~\citep{cryptoeprint:2026/103} develops a general hybrid protocol for encrypted regions of
interest (ROIs). These works establish that selective encryption is possible and can reduce
computation when only part of an input must be hidden. In this work, we characterize how encrypted dependencies propagate through
a network and when selective encryption retains a computational advantage. We notice that encrypting fewer input pixels does not
necessarily mean performing proportionally less homomorphic inference. Neural networks repeatedly
mix their inputs. After several layers, an activation far outside the original ROI may still depend on
an encrypted pixel and must therefore also be computed homomorphically.

This paper asks: \emph{when is selective encryption actually efficient?} We answer this through
\emph{encrypted-dependency propagation}, tracking which activations depend on encrypted inputs
and therefore must remain under FHE. For CNNs, this dependency region grows predictably with the
receptive field until saturation. We validate this behavior with real TFHE execution and use a
support-propagation predictor to estimate speedup for other architectures. Overall, efficiency is governed less by depth or model family than by how long
encrypted dependencies remain local.
\section{Method}
\label{sec:method}

Let $\mathbf{x}\in\mathbb{Z}^{n\times n}$ be an input and let
$R\subseteq[n]\times[n]$ denote the public index set of an $m\times m$ sensitive region of interest
(ROI), with $\tilde{\mathbf{x}}=\mathbf{x}[R]$. Full FHE encrypts all of $\mathbf{x}$ and evaluates
the complete circuit $\mathcal{C}_{\mathcal M}$. Instead, we encrypt only $\tilde{\mathbf{x}}$. Hence computations that depend only on the visible part of the input are
performed in plaintext, while computation influenced by the ROI $\tilde{x} $ remains encrypted. We call an
activation \emph{tainted} if it depends on at least one encrypted input value. Only tainted
activations require homomorphic evaluation, so the saving depends on how this region propagates
through the network.

\begin{figure}[t]
\centering
\includegraphics[width=1\linewidth]{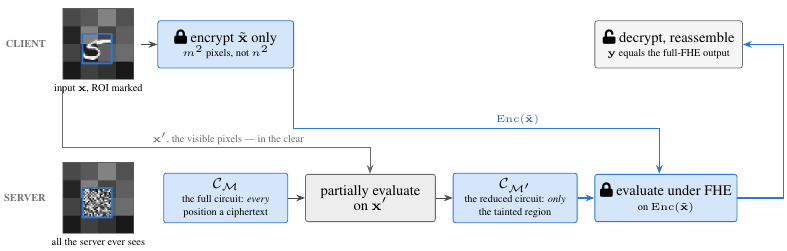}
\caption{Selective homomorphic inference evaluates only the computations that depend on the
encrypted ROI under FHE. The output remains identical to full encryption.}
\label{fig:method}
\end{figure}

Algorithm~\ref{alg:selective} and Figure~\ref{fig:method} summarize the procedure. We propagate the ROI to identify computations dependent on encrypted inputs, evaluate everything else in plaintext, and execute only the
remaining computation homomorphically while injecting plaintext-only contributions as clear constants. 

\begin{algorithm}[t]
\caption{\textsc{SelectiveHomomorphicEvaluation}}
\label{alg:selective}
\begin{algorithmic}[1]
\Require quantized network $\mathcal{M}$, input $\mathbf{x}$, public ROI index set $R$
\State \textbf{Client:} Send $R$ and the visible pixels
$x' = \mathbf{x}[R^c]$ to the server.
\State \textbf{Server:} Construct the full inference circuit $\mathcal{C}_\mathcal{M}$ for $\mathcal{M}$.
\State \textbf{Server:} Propagate $R$ through $\mathcal{C}_\mathcal{M}$ to identify the tainted computation.
\State \textbf{Server:} Partially evaluate $\mathcal{C}_\mathcal{M}$ on the visible pixels
$x'$ to get the reduced circuit $\mathcal{C}_{\mathcal{M}'}$.
\State \textbf{Client:} Encrypt $\tilde{\mathbf{x}} = \mathbf{x}[R]$ and send
$Enc(\tilde{\mathbf{x}})$ to the server.
\State \textbf{Server:} Evaluate $\mathcal{C}_{\mathcal{M}'}$ homomorphically on
$Enc(\tilde{\mathbf{x}})$ and return the encrypted output.
\State \textbf{Client:} Decrypt the encrypted output to obtain $\mathbf{y}$.
\end{algorithmic}
\end{algorithm}

The procedure is exact because each accumulator is simply split into encrypted and plaintext
contributions,
\begin{equation}
\sum_{j\in T} w_j a_j
+
\left(\sum_{j\notin T} w_j a_j+b\right)
=
\sum_j w_j a_j+b,
\label{eq:split}
\end{equation}
where \(T\) denotes the set of tainted input indices for the current accumulator. So, with the same fixed quantization and lookup tables, selective and full evaluation produce the
same output. Under the security of the underlying FHE scheme~\citep{10.1145/1536414.1536440, cryptoeprint:2018/421},
the encrypted ROI remains protected from the server, while the ROI index set and values outside the
ROI are intentionally public.

To analyze the efficiency of this method, we present the following simple lemma for convolutional networks:
\begin{lemma}[CNN locality and speedup]
\label{lem:speedup}
For a centered $m\times m$ encrypted ROI in an $n\times n$ input and a stack of stride-$1$,
valid $k\times k$ convolutions, the tainted region after layer $i$ has edge
\[
m_i=m+i(k-1)
\]
until saturation. When receptive-field growth is small relative to the input and ROI, the resulting
speedup scales as
\[
S=\Theta\!\left(\frac{n^2}{m^2}\right).
\]
\end{lemma}

The benefit lasts only while the tainted region remains smaller than the feature map. In this
idealized setting, it remains unsaturated through depth $d$ when
\begin{equation}
m+2d(k-1)<n.
\label{eq:validity}
\end{equation}
After saturation, later layers are effectively as expensive as full encryption. Thus, a small ROI
alone is insufficient: the architecture must preserve locality for enough of the network. The proof,
exact cost expression, and extensions to boundaries, stride, padding, and pooling are deferred to
Appendix~\ref{app:proof}.

\section{Experiments}
\label{sec:exp}

We evaluate channel-narrowed variants of AlexNet~\citep{NIPS2012_c399862d}, ResNet-18~\citep{he2015deepresiduallearningimage}, and
VGG-11~\citep{simonyan2015deepconvolutionalnetworkslargescale} that preserve each architecture's spatial dependency-propagation pattern at $224\times224$ using Concrete
TFHE~\citep{Concrete, cryptoeprint:2018/421} with 4-bit weights and activations. Full and selective circuits
share fixed quantization; timings are medians of three homomorphic evaluations excluding compilation,
key generation, encryption, and decryption. All selective circuits
reproduce the full-encryption integer output exactly.

\begin{figure}[t]
\centering
\includegraphics[width=0.95\linewidth]{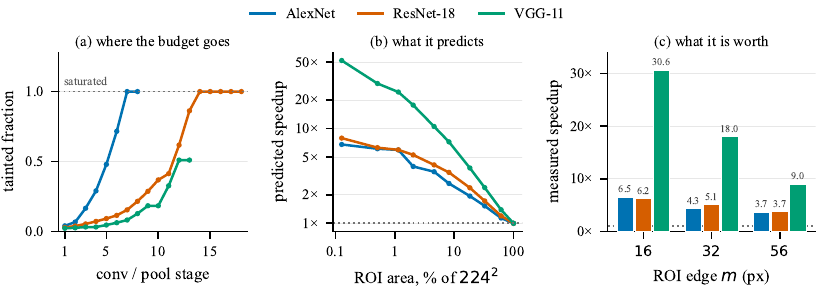}
\caption{The three panels connect dependency growth to realized speedup. (a) The tainted fraction
shows how quickly each architecture exhausts its locality budget across convolution and pooling
stages. (b) This propagation predicts how speedup changes as the encrypted ROI grows. (c) Real
TFHE measurements follow the same ordering: architectures that preserve locality longer achieve
larger gains.}
\label{fig:theory}
\end{figure}

As we see in Figure~\ref{fig:theory}, at a $32\times32$ ROI ($2.04\%$ of the input), VGG-11 achieves $18\times$ speedup, versus
$5.1\times$ for ResNet-18 and $4.3\times$ for AlexNet. The ordering follows early
down-sampling rather than depth: AlexNet's $11\times11$, stride-$4$ first convolution and
ResNet-18's early reductions consume locality much faster than VGG-11. More implementation and measurement details are deferred to Appendix~\ref{app:measurement}.

\subsection{Across architectures}

The results above suggest that architectural structure, rather than depth alone, determines
whether selective encryption is effective. We therefore extend the comparison to ConvNeXt-T~\citep{liu2022convnet2020s},
EfficientNet-B0~\citep{tan2020efficientnetrethinkingmodelscaling}, ViT-B/16~\citep{dosovitskiy2021imageworth16x16words},
Swin-T~\citep{liu2021swintransformerhierarchicalvision}, MLP-Mixer~\citep{tolstikhin2021mlpmixerallmlparchitecturevision}, and a fully connected
architecture.

\begin{figure}[t]
\centering
\includegraphics[width=0.82\linewidth]{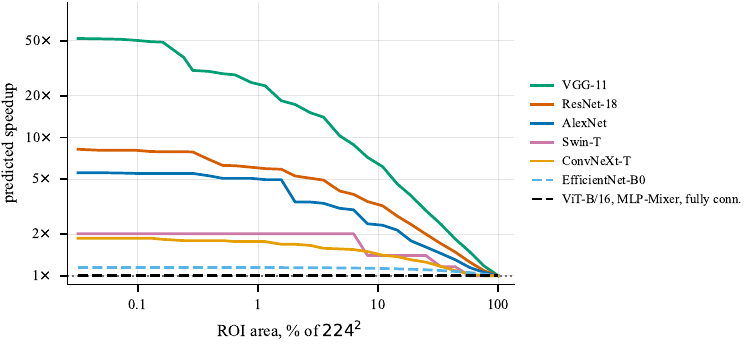}
\caption{Predicted speedup across architectures as the encrypted ROI grows. We estimate encrypted
computation from dependency propagation using
$S_{\mathrm{pred}}=\sum_a N_a/\sum_a T_a$, where $N_a$ and $T_a$ are the full and tainted
activation counts; full propagation rules are given in Appendix~\ref{app:predictor}.
Architectures that preserve local dependencies retain large gains, while those with early global mixing don't.}
\label{fig:results}
\end{figure}

Figure~\ref{fig:results} shows a clear separation according to how architectures mix spatial
information. At a $32\times32$ ROI, VGG-11 retains an $18\times$ speedup, whereas ViT,
MLP-Mixer, and a fully connected network don't benefit at all from this scheme because encrypted dependence
becomes global early. Swin-T retains $2.01\times$ through local windowed attention. Conversely,
convolution alone is not sufficient: EfficientNet-B0 reaches only $1.15\times$ because global
squeeze-and-excitation~\citep{hu2019squeezeandexcitationnetworks} spreads encrypted dependence across the feature map. The key architectural
property is therefore how long encrypted dependencies remain local.
\section{Discussion}
\label{sec:discussion}

Our results suggest treating \emph{encrypted-dependency propagation} as an explicit architectural
criterion for privacy-preserving neural networks. Alongside arithmetic complexity and nonlinear
operations, choices such as kernel size, stride, pooling, attention windows, and global reductions
determine how quickly a network consumes its ``locality budget.'' The support-propagation predictor
could therefore serve as a design or architecture-search objective, favoring models that keep
sensitive dependencies local for longer.

Selective inference also introduces a privacy--efficiency tradeoff. Smaller encrypted regions reduce
homomorphic work, but only values inside the ROI are cryptographically protected. The relevant
operating point is therefore the smallest ROI that covers the sensitive content, and ROI size should
be considered jointly with the architecture's dependency propagation.

\section{Limitations}
The predictor is an architectural cost model, not an exact runtime simulator. Thus, non-CNN results (Figure \ref{fig:results}) are predictions rather than measured FHE runtimes. The wall-clock experiments use channel-narrowed variants of AlexNet, ResNet-18, and VGG-11 to make FHE evaluation tractable. These variants preserve the spatial dependency-propagation pattern, but their absolute runtimes should not be interpreted as those of the original full-width networks.

\section*{Acknowledgements}
This work began as a final project for MIT 6.5610 (Spring 2025), and an earlier version of it
appeared as the non-archival course report \emph{Encrypt What Matters: ROI-Guided FHE for CNN
Inference}. We thank the 6.5610 course
staff for their feedback on the earlier version.

\bibliographystyle{plainnat}
\bibliography{refs}

\appendix
\section{Measurement Details}
\label{app:measurement}

This appendix provides the experimental configuration used for the wall-clock measurements in
Section~\ref{sec:exp}. We report both absolute homomorphic-evaluation latencies and the corresponding
speedups.

\paragraph{Hardware and software.}
All experiments were run on a single Apple M1 Pro (MacBookPro18,1) with 10 CPU cores
(8 performance and 2 efficiency cores) and 16\,GiB of unified memory, running macOS~26.6.2
(arm64). Homomorphic evaluation was CPU-only; no GPU acceleration was used. The software
environment consisted of Python~3.11.16, \texttt{concrete-python}~2.11.0, and NumPy~1.26.4.
Both the full and selective circuits were evaluated with the same thread configuration,
using 8 threads via \texttt{OMP\_NUM\_THREADS=8} and
\texttt{RAYON\_NUM\_THREADS=8}.

\paragraph{Cryptographic parameters.}
TFHE parameters were selected independently for each compiled circuit by the Concrete optimizer
using the multi-parameter strategy (\texttt{MULTI}, with
\texttt{single\_precision=False}). We retained Concrete's default 128-bit security level and
default error target. Because parameter selection is performed separately for each circuit,
the optimizer may choose less expensive parameters for a selective circuit when its arithmetic
requirements are smaller. This effect is therefore included in the measured wall-clock speedup.

Across the circuits used in our experiments, the compiler-reported error probabilities ranged
from $5.4\times10^{-11}$ to $1.3\times10^{-10}$ for the full circuits and from
$3.4\times10^{-10}$ to $4.8\times10^{-10}$ for the selective circuits.

\paragraph{Quantization and programmable bootstrapping.}
Activations are represented as 4-bit unsigned integers. Weights are quantized to 4-bit symmetric
signed integers in $[-7,7]$, using a per-tensor scale of
\[
    s_w = \frac{\max |w|}{7}.
\]
Each convolutional or dense layer first computes an integer accumulator and then applies a single
lookup table implemented by programmable bootstrapping (PBS). The lookup table jointly performs
the nonlinearity and output re-quantization.

For each layer, we derive a provable lower bound on the accumulator from the quantized weights and
the input range. The accumulator is offset by this bound and rounded so that the resulting lookup-table
index fits within 6 bits. Thus, the programmable bootstrap operates on a 6-bit table index even when
the unrounded integer accumulator has a larger range.

Output scales are calibrated using the 99.9th percentile of positive accumulator values over
256 calibration images. The selective circuit reuses the same lookup tables, accumulator
bounds, rounding parameters, and output scales as the corresponding full circuit. Partial evaluation
only separates plaintext and encrypted contributions to the same integer accumulator; consequently,
the accumulator presented to each lookup table is identical in the full and selective executions.

For the timing experiments, network weights were trained in floating point on an MNIST-on-canvas
dataset containing 12,000 training images and 2,000 test images for 25 epochs, then quantized and
frozen before FHE compilation.

\paragraph{Channel width.}
The encrypted-dependency mask is spatial: all channels at a tainted spatial location are treated
identically. To make the wall-clock FHE experiments tractable, we therefore use channel-narrowed
versions of the three convolutional architectures: AlexNet at $1/16$ of its original channel width,
ResNet-18 at $1/32$, and VGG-11 at $1/64$. Each scaling factor is an exact divisor of the channel
counts in the corresponding architecture.

Channel narrowing does not change the spatial dependency-propagation pattern, and both the full and
selective circuits use the same narrowed architecture in each comparison. The absolute latencies and
measured speedups reported in Table~\ref{tab:latency} therefore refer to these narrowed variants;
they should not be interpreted as absolute runtimes of the original full-width networks.

\paragraph{Timing protocol.}
Evaluation latency is the median of three executions of \texttt{circuit.run} on a single
$224\times224$ input. The reported evaluation time excludes compilation, key generation,
encryption, and decryption. These costs are reported separately in
Table~\ref{tab:overhead}.

Each circuit was compiled and keyed once and then reused for all three timed executions. For each
architecture, the full-encryption baseline was measured once (as the median of three executions)
and reused when computing the speedup for each ROI size.

Before collecting timing measurements, we verified correctness against the plaintext quantized
implementation. Full and selective executions produced identical integer outputs on eight held-out
inputs under Concrete simulation. We additionally verified under real TFHE keys that the decrypted
outputs of both circuits matched the plaintext reference. All configurations reported below passed
these checks.

\begin{table}[t]
    \centering
    \footnotesize
    \caption{
        Homomorphic-evaluation latency for $224\times224$ inputs.
        Times are medians of three runs, with interquartile ranges in parentheses.
        ``PBS'' denotes the number of compiled programmable bootstraps.
        The full-encryption baseline is measured once per architecture and shared across ROI sizes.
    }
    \label{tab:latency}
    \setlength{\tabcolsep}{3.5pt}
    \begin{tabular}{llrrrrrrr}
        \toprule
        Architecture & Width & $m$ & ROI &
        PBS (full) & PBS (sel.) &
        Full (s) & Selective (s) & Speedup \\
        \midrule

        AlexNet
        & $1/16$ & $16$ & $0.51\%$
        & $356{,}480$ & $58{,}112$
        & $1571.7\ (43.3)$ & $241.6\ (0.8)$
        & $6.51\times$ \\

        &
        & $32$ & $2.04\%$
        & & $89{,}472$
        & & $362.2\ (2.1)$
        & $4.34\times$ \\

        &
        & $56$ & $6.25\%$
        & & $105{,}216$
        & & $426.8\ (9.9)$
        & $3.68\times$ \\

        \midrule

        ResNet-18
        & $1/32$ & $16$ & $0.51\%$
        & $485{,}296$ & $81{,}140$
        & $2166.2\ (4.3)$ & $349.4\ (0.6)$
        & $6.20\times$ \\

        &
        & $32$ & $2.04\%$
        & & $95{,}620$
        & & $421.5\ (9.5)$
        & $5.14\times$ \\

        &
        & $56$ & $6.25\%$
        & & $128{,}172$
        & & $577.8\ (39.9)$
        & $3.75\times$ \\

        \midrule

        VGG-11
        & $1/64$ & $16$ & $0.51\%$
        & $644{,}448$ & $26{,}864$
        & $4517.7\ (38.2)$ & $147.5\ (0.9)$
        & $30.63\times$ \\

        &
        & $32$ & $2.04\%$
        & & $42{,}896$
        & & $251.1\ (0.5)$
        & $17.99\times$ \\

        &
        & $56$ & $6.25\%$
        & & $80{,}368$
        & & $503.7\ (0.6)$
        & $8.97\times$ \\

        \bottomrule
    \end{tabular}
\end{table}

At the $32\times32$ ROI used in the main comparison, for example, the measured VGG-11
homomorphic-evaluation time decreases from $4517.7$\,s under full encryption to $251.1$\,s
under selective evaluation, corresponding to a $17.99\times$ speedup.

\begin{table}[t]
    \centering
    \footnotesize
    \caption{
        Costs excluded from the homomorphic-evaluation measurements in
        Table~\ref{tab:latency}, reported at $m=32$.
        Compilation and key generation are performed once per circuit and can be reused across
        inference requests. Partial evaluation is performed once per input.
        Decryption takes less than $0.02$\,s in every configuration and is omitted.
    }
    \label{tab:overhead}
    \setlength{\tabcolsep}{4pt}
    \begin{tabular}{lrrrrrrr}
        \toprule
        & \multicolumn{2}{c}{Compile (s)}
        & \multicolumn{2}{c}{Key generation (s)}
        & \multicolumn{2}{c}{Encryption (s)}
        & Partial eval. (s) \\
        \cmidrule(lr){2-3}
        \cmidrule(lr){4-5}
        \cmidrule(lr){6-7}
        Architecture
        & Full & Sel.
        & Full & Sel.
        & Full & Sel.
        & Sel. \\
        \midrule

        AlexNet
        & $3014.0$ & $2936.7$
        & $15.8$ & $13.5$
        & $5.44$ & $0.11$
        & $0.007$ \\

        ResNet-18
        & $2782.3$ & $2886.5$
        & $13.8$ & $14.6$
        & $3.92$ & $0.09$
        & $0.007$ \\

        VGG-11
        & $178.9$ & $184.3$
        & $18.9$ & $17.2$
        & $3.36$ & $0.08$
        & $0.004$ \\

        \bottomrule
    \end{tabular}
\end{table}

The additional per-input cost introduced by selective evaluation is therefore negligible relative
to homomorphic evaluation. At $m=32$, the reduction in homomorphic-evaluation time is substantially
larger than the difference in one-time setup cost for all three measured architectures.

\section{CNN Locality and Cost Scaling}
\label{app:proof}

We now derive the locality result used in Lemma~\ref{lem:speedup}. The closed-form analysis
considers a centered $m\times m$ encrypted ROI in an $n\times n$ input and a stack of stride-$1$,
valid $k\times k$ convolutions. More general architectural operations are handled by the
support-propagation procedure in Appendix~\ref{app:predictor}.

Consider first one spatial dimension. Let a tainted interval have length $\ell$. An output of a
stride-$1$ convolution is tainted exactly when its receptive field intersects this interval.
Provided the expanding interval has not reached a feature-map boundary, a convolution with kernel
width $k$ therefore increases its length to
\[
\ell+k-1.
\]
Applying the same argument independently along both spatial axes, the tainted region after $i$
layers is a square with edge
\begin{equation}
m+i(k-1).
\label{eq:tainted-growth}
\end{equation}

At the same time, each valid convolution reduces the feature-map edge. After $i$ layers,
\begin{equation}
n_i=n-i(k-1).
\label{eq:feature-growth}
\end{equation}
Consequently, the exact tainted edge for a centered ROI is
\begin{equation}
m_i
=
\min\!\left(
m+i(k-1),\,
n-i(k-1)
\right).
\label{eq:exact-tainted-edge}
\end{equation}
Before saturation, the first term is smaller and
Equation~\eqref{eq:exact-tainted-edge} reduces to
Equation~\eqref{eq:tainted-growth}.

Saturation occurs when the expanding tainted region reaches the boundary of the shrinking feature
map. Requiring the ROI to remain unsaturated through depth $d$ gives
\begin{equation}
m+d(k-1)
<
n-d(k-1),
\end{equation}
or equivalently
\begin{equation}
m+2d(k-1)<n,
\end{equation}
which is Equation~\eqref{eq:validity} in the main text.

For the same idealized stack, assume that encrypted computation has constant cost per spatial
position. Before saturation, the selective cost is
\begin{equation}
C_{\mathrm{ROI}}
=
\Theta\!\left(
\sum_{i=1}^{d}
[m+i(k-1)]^2
\right),
\end{equation}
whereas full encryption evaluates every spatial position and therefore costs
\begin{equation}
C_{\mathrm{full}}
=
\Theta\!\left(
\sum_{i=1}^{d}
[n-i(k-1)]^2
\right).
\end{equation}
The resulting speedup is
\begin{equation}
S
=
\Theta\!\left(
\frac{
\sum_{i=1}^{d}[n-i(k-1)]^2
}{
\sum_{i=1}^{d}[m+i(k-1)]^2
}
\right).
\label{eq:exact-speedup}
\end{equation}

When receptive-field growth is small relative to both $m$ and $n$, the numerator and denominator
scale respectively as $dn^2$ and $dm^2$. Hence
\begin{equation}
S
=
\Theta\!\left(
\frac{n^2}{m^2}
\right),
\end{equation}
which is the $\Theta(n^2/m^2)$ scaling stated in
Lemma~\ref{lem:speedup}.

The centered-ROI assumption gives the cleanest closed form. Near an image boundary, dependency
growth can be clipped because part of the receptive field lies outside the valid feature map.
Similarly, stride, padding, pooling, residual connections, attention, and other architectural
operations change dependency propagation in ways that are more naturally handled directly on the
network graph. We therefore use the numerical support-propagation procedure below for the
architectures evaluated in the experiments.

\section{Support-Propagation Predictor and Limitations}
\label{app:predictor}

The predictor replaces the numerical forward pass with binary dependency propagation. The encrypted
ROI initializes the support mask, and an activation is marked tainted whenever it can depend on an
encrypted input. Learned weights are treated as structurally active, while additive constants do not
introduce dependency.

Propagation follows the architecture. Convolution and pooling taint an output when its receptive
field intersects the tainted region; point-wise operations preserve support; and additions or residual
connections take the union of their input supports. Local attention propagates support within its
window, as in Swin Transformer~\citep{liu2021swintransformerhierarchicalvision}, whereas architectures with global token mixing,
such as ViT and MLP-Mixer~\citep{dosovitskiy2021imageworth16x16words, tolstikhin2021mlpmixerallmlparchitecturevision}, can make dependency
global once any participating input is tainted. Global reductions are handled similarly, including
squeeze-and-excitation~\citep{hu2019squeezeandexcitationnetworks}, where a tainted reduced value can subsequently be
broadcast across the feature map.

For each recorded activation $a$, let $N_a$ denote the total number of activation elements and
$T_a$ the number marked tainted. The predicted speedup is
\begin{equation}
S_{\mathrm{pred}}
=
\frac{\sum_a N_a}{\sum_a T_a}.
\label{eq:predicted-speedup}
\end{equation}

\begin{algorithm}[h]
\caption{\textsc{PredictSpeedup}}
\label{alg:support}
\begin{algorithmic}[1]
\Require architecture $\mathcal M$, encrypted-input mask $R$
\State Propagate an all-ones mask through $\mathcal M$ and record $N_a$ at each activation.
\State Propagate $R$ using the dependency rules above and record the tainted count $T_a$.
\State \Return $\displaystyle S_{\mathrm{pred}}=\frac{\sum_a N_a}{\sum_a T_a}$.
\end{algorithmic}
\end{algorithm}

We check that an empty ROI produces no tainted computation, a whole-image ROI gives exactly
$1.00\times$ predicted speedup, and tainted computation is non-decreasing with ROI size. Across the
nine configurations with real TFHE measurements, the predictor has $2.7\%$ median error and
$8.3\%$ worst-case error.

\end{document}